\pdfoutput=1

\documentclass[a4paper,fleqn]{cas-dc}

\usepackage[numbers]{natbib}

\def\S{\emph{S}}
\def\SS{\emph{Sq}}
\def\CT{\emph{c}}
\def\CD{\emph{D}}

\def\AVG{\emph{$\bar{p}$}}
\def\COV{\emph{C}}

\usepackage[ruled,vlined]{algorithm2e}
\usepackage{amsmath}
\usepackage{subcaption}
\usepackage[font=small,skip=7pt]{caption}

\def\tsc#1{\csdef{#1}{\textsc{\lowercase{#1}}\xspace}}
\tsc{WGM}
\tsc{QE}
\begin{document}
\let\WriteBookmarks\relax
\def\floatpagepagefraction{1}
\def\textpagefraction{.001}


\shorttitle{Feature-assisted interactive geometry reconstruction in 3D point clouds using incremental region growing}    
\shortauthors{Szabo et al.}  
\title [mode = title]{Feature-assisted interactive geometry reconstruction in 3D point clouds using incremental region growing}  

\author[1]{Attila Szabo}[orcid=0000-0002-1322-8704,twitter=AszaboKing]
\cormark[1]
\fnmark[1]
\ead{aszabo@vrvis.at}
\credit{Conceptualization, Methodology, Software, Visualization, Writing - Original Draft, Writing - Review \& Editing}

\author[1]{Georg Haaser}
\ead{haaser@vrvis.at}
\credit{Conceptualization, Software, Methodology, Validation}

\author[1]{Harald Steinlechner}
\ead{steinlechner@vrvis.at}
\credit{Investigation, Validation, Formal analysis, Writing - Original Draft, Writing - Review \& Editing}

\author[1]{Andreas Walch}[orcid=0000-0002-4567-7942]
\ead{walch@vrvis.at}
\credit{Writing - Original Draft}

\author[1]{Stefan Maierhofer}
\ead{maierhofer@vrvis.at}
\credit{Supervision, Funding acquisition}

\author[1]{Thomas Ortner}[orcid=0000-0002-9373-6409]
\ead{inbox@ortner.fyi}
\ead[url]{http://ortner.fyi/}
\credit{Writing - Original Draft, Writing - Review \& Editing, Project administration}

\author[2]{Eduard Gröller}
\ead{groeller@cg.tuwien.ac.at}
\credit{Supervision, Project administration}

\affiliation[1]{organization={VRVis Zentrum f\"{u}r Virtual Reality und Visualisierung Forschungs-GmbH},
	addressline={Donau-City-Straße 11}, 
	city={Vienna},
	postcode={1220}, 
	country={Austria}}

\affiliation[2]{organization={TU Wien},
	addressline={Karlsplatz 13}, 
	city={Vienna},
	postcode={1040}, 
	country={Austria}}



\begin{abstract}
Reconstructing geometric shapes from point clouds is a common task that is often accomplished by experts manually modeling geometries in CAD-capable software. 
State-of-the-art workflows based on fully automatic geometry extraction are limited by point cloud density and memory constraints, and require pre- and post-processing by the user.
In this work, we present a framework for interactive, user-driven, feature-assisted geometry reconstruction from arbitrarily sized point clouds. 
Based on seeded region-growing point cloud segmentation, the user interactively extracts planar pieces of geometry and utilizes contextual suggestions to point out plane surfaces, normal and tangential directions, and edges and corners. 
We implement a set of feature-assisted tools for high-precision modeling tasks in architecture and urban surveying scenarios, enabling instant-feedback interactive point cloud manipulation on large-scale data collected from real-world building interiors and facades. 
We evaluate our results through systematic measurement of the reconstruction accuracy, and interviews with domain experts who deploy our framework in a commercial setting and give both structured and subjective feedback.
\end{abstract}

\begin{graphicalabstract}
\includegraphics{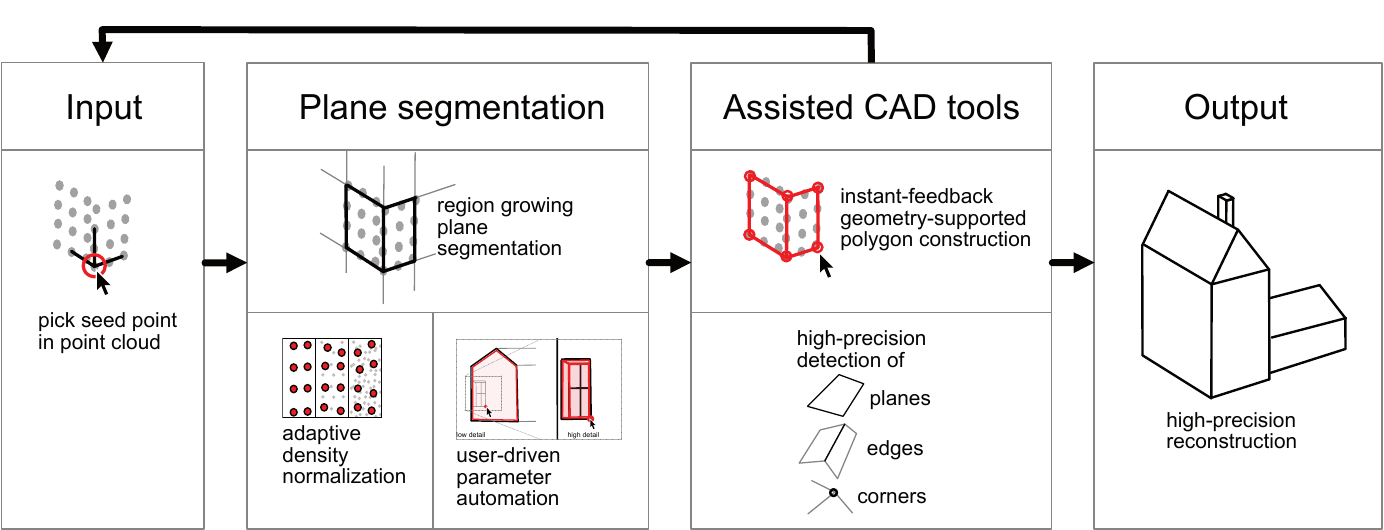}
\end{graphicalabstract}

\begin{highlights}
\item Real-world point clouds show strong heterogeneity in size, density, and quality
\item Fully automated geometry reconstruction almost always requires human intervention or quality control
\item Human-in-the-loop approach avoids cumbersome filter-and-repair post-processing
\item Guided technique effectively utilizes human intent to navigate difficult reconstruction scenarios
\end{highlights}



\maketitle


\section{Introduction}
\label{sec:introduction}

\begin{figure*}
  \includegraphics[width=\linewidth]{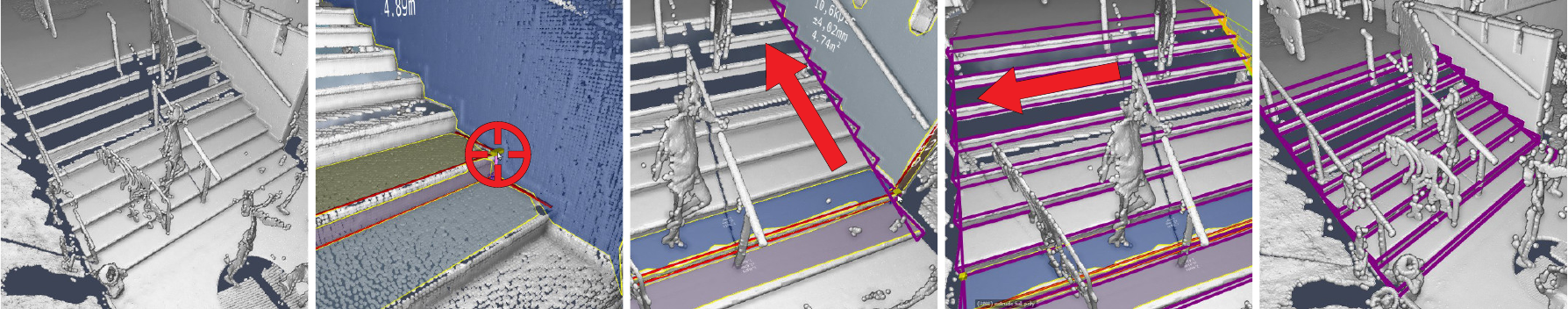}
  \centering
   \caption{Reconstructing stairs in a real-world point cloud. From left to right, the human operator first zooms in on the stairs and reconstructs the side of a step in detail. Then they zoom out and copy the reconstructed geometry into where points are missing. Finally, they extrude the stair geometry along the tangential direction of the step's face. The reconstruction of the staircase's shape is then complete.}
 \label{fig:teaser}
 \end{figure*}

A major goal in the field of surveying and mapping is to create Computer Aided Design (CAD)-ready geometrical models that accurately describe the as-built conditions of buildings’ inside and outside structures. It is important to represent walls, but also more intricate features, such as, roofs, window sills, or a flight of stairs, as shown in Figure \ref{fig:teaser}.

As a means to this end, surveyors capture real-world buildings with terrestrial laser scanners producing 3D point clouds. 
Depending on the size of the building these typically range from tens of millions to billions of points. 
This high-detail representation is bulky and impractical for CAD workflows, construction documentation, and as-designed comparisons. 
To create abstracted, CAD-ready models, for instance an accurately measured 3D plan, surveyors strive to derive edges and corners from point clouds.

\begin{figure*}[b]
  \centering
  \begin{subfigure}[]{0.46\linewidth}
    \centering
    \includegraphics[width=\linewidth]{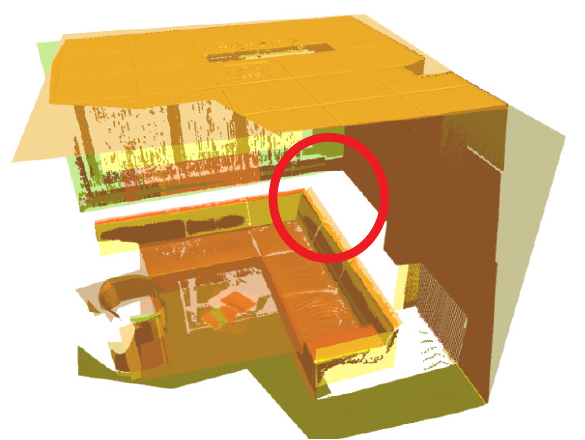}
    \label{subfig:errorsinpointcloudsc}
  \end{subfigure}
  \begin{subfigure}[]{0.46\linewidth}
    \centering
    \includegraphics[width=\linewidth]{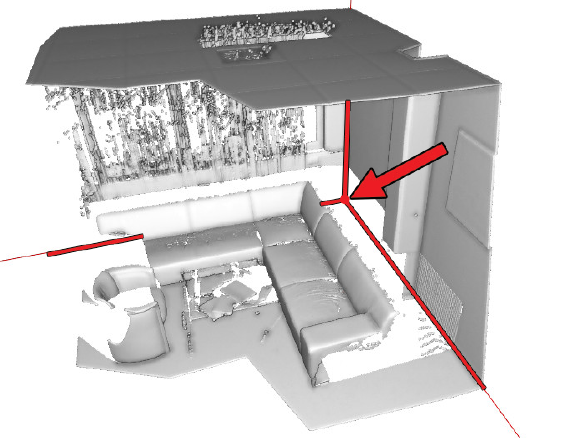}
    \label{subfig:errorsinpointcloudsd}
  \end{subfigure}
  \caption{Left: Occlusions cause holes in the automatic segmentation (yellow). The desired corner is missing (red circle). Right: The corner of interest (red arrow) is at the intersection of the walls and the floor.}
  \label{fig:problemholes}
\end{figure*}

Several properties of real-world laser scans make feature derivation challenging: relevant edges and corners are not well captured due to either sampling constraints or scanning occlusions. Further, real-world laser scans contain unwanted points, captured from vegetation, furniture, appliances, or people walking through the scan (can be seen in Figure \ref{fig:teaser}). Finally, high-precision scanning leads to large point clouds generally requiring an out-of-core approach.

Since a majority of human-made structures can be described as piece-wise planar objects, off-the-shelf tools typically allow users to globally fit plane primitives~\cite{schnabel2007efficient} to a point cloud, so the resulting planes may act as shape proxies to support reconstruction in orthographic views. 
Automatic algorithms often incur over- or undersegmentation due to global scope and parametrization. This requires prior cleaning and subsampling, as well as post-processing to exclude unwanted results and to manually construct missing results. 
Previous approaches typically do not go far enough in supporting the human operator in these tasks, which we aim to rectify in this work.
Our system also focuses on buildings consisting of planar structures. 
More complex support shapes are considered future work, which is discussed in Section \ref{sec:discussion}.

\begin{figure*}[h]
  \centering
  \begin{subfigure}[]{.46\linewidth}
    \centering
    \includegraphics[width=\linewidth]{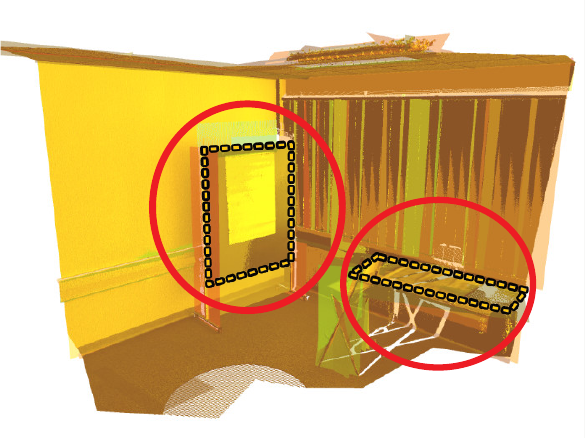}
    \label{subfig:errorsinpointcloudsc}
  \end{subfigure}
  \begin{subfigure}[]{.46\linewidth}
    \centering
    \includegraphics[width=\linewidth]{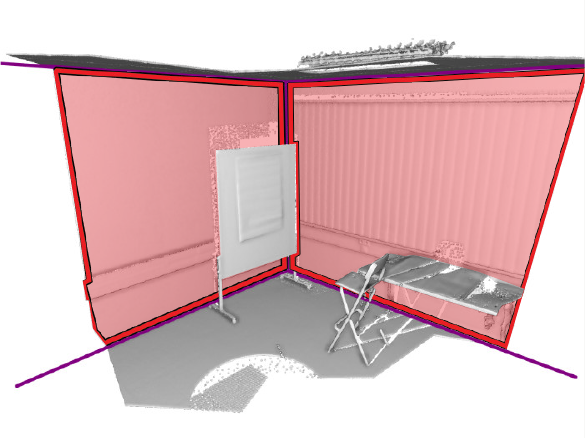}
    \label{subfig:errorsinpointcloudsd}
  \end{subfigure}
  \caption{Left: Automatic segmentation (yellow) contains unwanted segments, e.g. whiteboard or desk (red circles). Right: Segments of interest are the two walls (red outlines).}
  \label{fig:problemtoomany}
\end{figure*}

We propose a novel interaction-based framework that does not require any global pre-segmentation or point cleaning steps (Figures \ref{fig:problemholes},\ref{fig:problemtoomany} left), while offering users a constructive approach to reconstructing corners and edges despite the presence of missing data and unwanted points (Figures \ref{fig:problemholes},\ref{fig:problemtoomany} right). 
Our human-in-the-loop approach provides algorithmic assistance to the user through on-demand region-growing plane-segmentation with instant feedback enabled by an adaptive point cloud resolution scheme and robust edge and corner detection. 
Region growing as well as primitive fitting operate \emph{locally} and are evaluated \emph{incrementally}. 

\begin{figure*}[b]
  \includegraphics[width=\linewidth]{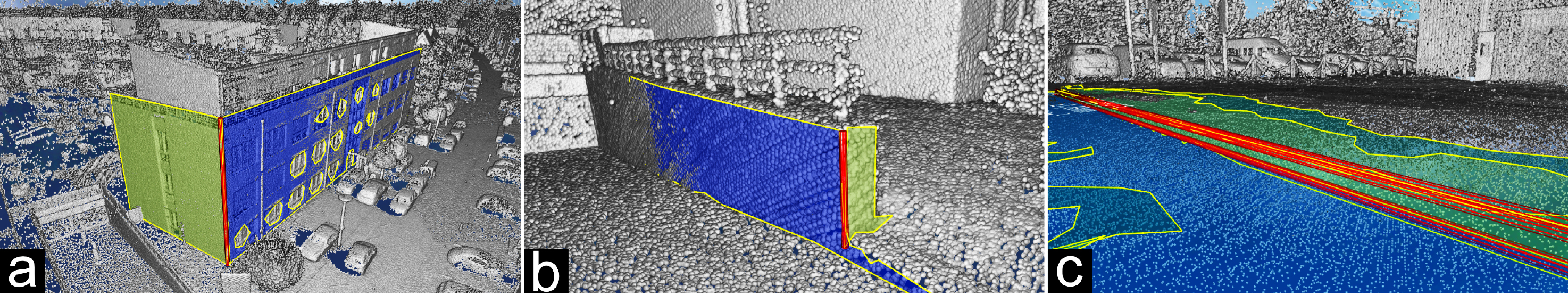}
  \caption{Interactive reconstruction at different degrees of detail. Cloud has 1 billion points. (a) large building ($\sim$ 20m long). (b) sidewalk structure ($\sim$ 1m high). (c) roadside curb ($\sim$ 0.15m high).}
  \label{fig:reconstructionexamples}
\end{figure*}

Our work is an extension of Steinlechner et al.~\cite{steinlechner2019adaptive}. There the authors fit plane segments to point clouds and support user interactions, for instance, to prevent pick-through when measuring distances. 
In this work we use a similar out-of-core-strategy. 
Our technique supports a full range of tools for reconstruction at any detail level, as seen in Figure~\ref{fig:reconstructionexamples}.
Existing interactive reconstruction-based tools require a global pre-segmentation of the point cloud~\cite{lejemble2020persistence},~\cite{arikan2013snap}.
Our approach is orthogonal as all segmentation is local, on-demand, and user-driven with a tightly integrated interaction loop. 

An overview of our system is shown in Figure \ref{fig:graphicalabstract}.

\subsection{Contributions}
In summary, we present the following contributions:
\begin{itemize}
\item A robust on-the-fly density normalization scheme permitting level-of-detail reconstruction of arbitrary-sized out-of-core point clouds (Section \ref{subsec:adaptiveresolution}), 
\item A novel point cloud interaction framework integrated with on-demand incremental plane segmentation (Sections \ref{subsec:regiongrowingloop}), and
\item A user-driven workflow for feature-assisted high-precision geometry reconstruction in non-preprocessed point clouds (Sections \ref{sec:assistedgeometryconstructiontools}).
\end{itemize}

\subsection{Structure of the Paper}
Section \ref{sec:relatedwork} reviews recent works in the field of interactive point cloud geometry reconstruction. Sections \ref{sec:pointcloudsegmentation} and \ref{sec:assistedgeometryconstructiontools} detail our incremental region growing and robust interaction loop, respectively. 
In Section \ref{sec:evaluation} we evaluate our three contributions with synthetic tests, example showcases, and real-world expert user interviews, respectively.
Section \ref{sec:discussion} discusses strengths and shortcomings of our approach.

\section{Related Work}
\label{sec:relatedwork}

\begin{figure*}
  \centering
  \includegraphics[width=0.875\linewidth]{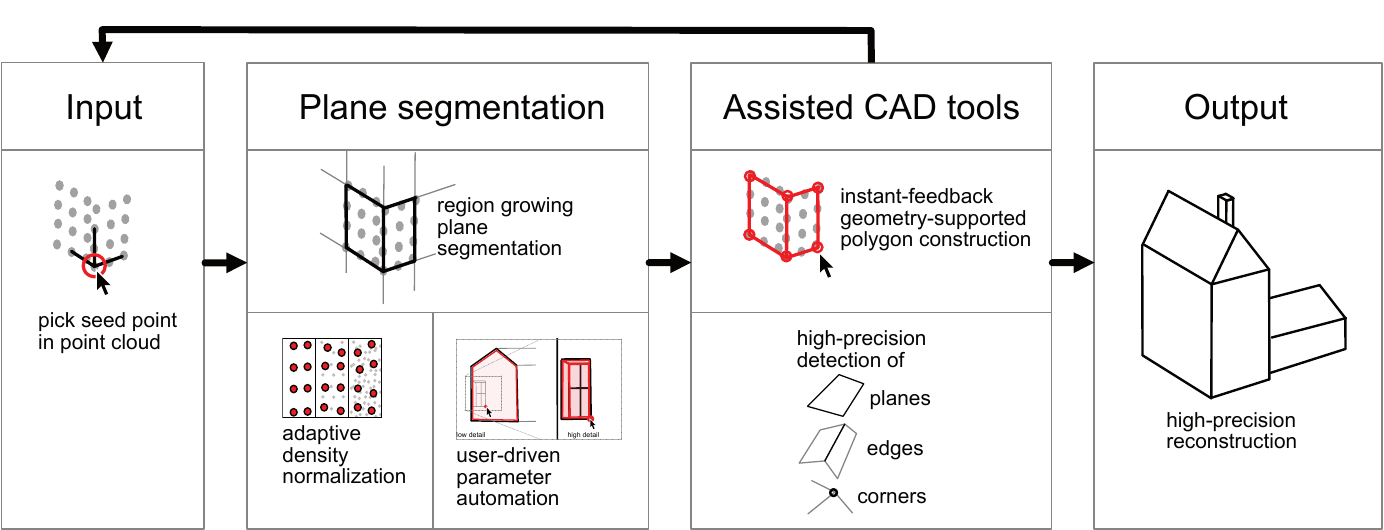}
  \caption{System overview. From left to right, the four labeled boxes represent the major components of our approach. \textit{Input}: In a point cloud a seed point is picked by the human operator from a specific view. \textit{Plane segmentation}: The point cloud is locally segmented into planar regions starting from the seed point. Segmentation parameters are automatically estimated from the user's view. Our adaptive density normalization scheme ensures consistent convergence across regions of too low or too high density. \textit{Assisted CAD tools}: High-precision planes, edges and corners are synthesized from planar regions and previously-reconstructed geometries, which assist the user in drawing the real shape represented by the point cloud. These support geometries are available to the user instantly after picking the seed point, and progressively get more refined in subsequent iterations. \textit{Output}: Repeated reconstruction of different parts of the point cloud yields the desired final high-precision geometry.}
  \label{fig:graphicalabstract}
\end{figure*}

We structure the body of related work into two sections, automatic methods (Section \ref{subsec:related_automatic}), and user-driven (Section \ref{subsec:related_userdriven}) methods that ingest point clouds with real-world artifacts capturing piece-wise planar objects. 
For an extensive review of surface reconstruction from, and geometric primitives detection in, point clouds, we refer to Berger et al.~\cite{berger2017survey} and Kaiser et al.~\cite{kaiser2019survey}, respectively.

\subsection{Automatic Methods}
\label{subsec:related_automatic}

Automatic methods have a longer history of research than user-driven ones. Common approaches attempt to reconstruct a plausible surface model, often planes, either through RANSAC~\cite{schnabel2007efficient} or region growing~\cite{rabbani2006segmentation}. Both are available in off-the-shelf tools~\cite{Rusu_ICRA2011_PCL, cgal:ovja-pssd-21b}. RANSAC requires careful parameter tuning to achieve the desired level of detail representation and region growing is sensitive to initial conditions, noise, and outliers~\cite{kaiser2019survey}. 

Tuning automatic shape detection represents a trade-off between missing important geometric detail versus receiving a large number of primitives, which requires manual cleanup. More advanced methods detect planes and then automatically prune or merge them according to different criteria, such as principal orientations, parallelism, orthogonality, or coplanarity~\cite{li2011globfit, monszpart2015rapter, oesau2016planar, nan2017polyfit}. Since processing a large number of detected shapes is computationally intensive, Bauchet et al.~\cite{bauchet2020kinetic} employ a kinetic polyhedral approach after shape detection that performs well for large amounts of measurement data.

\subsubsection{Level of Detail}

Level-of-detail-aware approaches employ structured metadata to cut down on operations or memory consumption. 
Real-time point-cloud renderers typically use octree-based data structures to subdivide space into cells of equal point density.
They dynamically fill the limited graphics memory with a uniformly distributed subset of points to guarantee optimal image quality.
Region-growing segmentation techniques in particular utilize spatial indexing schemes to group points, typically voxel grids or octree-equivalents. 
Deschaud et al.~\cite{deschaud2010fast} decompose the point cloud into a voxel grid as a means of batch-processing the region growing algorithm and avoiding costly point-neighborhood searches. 
Vo et al.~\cite{vo2015octree} use an octree based on point planarity rather than density. They sidestep the need for point operations almost entirely by comparing representative plane segments instead.
For our work we chose a region-growing algorithm equivalent to the implementation in the Point Cloud Library~\cite{Rusu_ICRA2011_PCL}. Extending it with our ad-hoc density normalization scheme does not require a specifically crafted data structure, thus enabling a minimal-effort implementation into existing point renderers.

More recently, Mercier et al.~\cite{mercier2022moving} present a level-of-detail-aware reconstruction technique for algebraic surfaces. It uses an octree to store intermediate surface computations and an adaptive traversal scheme to obtain appropriate resolutions for different regions. 
Their octree traversal is in principle comparable to our adaptive resolution selection (Section \ref{subsec:adaptiveresolution}), and could be adapted for planar surfaces. 

\subsubsection{Feature Detection}

2D feature detectors are long-standing image processing steps, including the Canny edge detector and Harris corner detector. They do not generalize straightforwardly, since point clouds exhibit highly irregular sampling patterns and scan artifacts, as compared to regular pixel grids, serving as motivation for much research.

Filter-based approaches exist in 3D, and overcome the aforementioned problems by applying rule-based constraints (e.g. merge close line segments, make lines parallel/orthogonal). 
Hackel et al.~\cite{hackel2016contour} use classifiers to assign points to contour lines and select consistent contour sets as plausible complete outlines of structures.
More recently, Wang et al.~\cite{wang2020pie} use deep learning to infer the parameters of contour curves from classified points. 
Himeur et al.~\cite{himeur2021pcednet} further distinguish edge types (hard/smooth). They provide an interactive element for user-specified edge definitions, being able to handle compromised real-world point clouds.
We, in comparison, specifically understand edges as being delineated by large planes, which are commonly used in building planning.
Even more sophisticated feature extraction techniques could be used in our system if required by the application. 

In contrast to \emph{global} optimization methods, we propose an efficient \emph{local} technique for reconstructing planes, edges, and corners on-demand in a limited spatial region around a user-picked seed point.
Our workflow's constructive nature avoids any prior processing (beyond octree construction), removal of artifacts, time-consuming parameter tuning, or manual clean-up of segmentation results. 
Furthermore, the user is given a set of tools to deal with noise, missing points, and unwanted portions of a scan based on human decision-making and contextual awareness. 
Since our interactive approach is tightly coupled with point rendering, the density-based octree re-uses existing infrastructure. 
Limiting reconstruction space makes the process independent of point cloud size while retaining full-detail accuracy.

\subsection{User-driven Methods}
\label{subsec:related_userdriven}

Commercial point processing tools, such as ArcGIS or AutoCAD, include user-driven surface reconstruction tailored to the domain of CAD. Workflows consist of a non-interactive and an interactive part, where a global algorithm first finds dominant planes in the subsampled point cloud automatically. Subsequently, the user iterates through orthographic projections on planes of interest. Polygonal geometry is built in mixed 2D and 3D views using classic construction tools. In comparison, our approach lets the user manipulate the point cloud directly. This avoids the loss of locality caused by multiple views and leverages contextual awareness and semantic information provided by the three dimensional viewpoint. 

Chen and Chen~\cite{chen2008architectural} recover planar segments and their connectivity through edges and corners automatically. They assist the user in repairing holes through topological reasoning on regularity and symmetry of the polygons. This approach combines multiple global processing steps, with user interaction at the end to complete the result. Our proposed on-demand reconstruction does not require global pre-segmentation and incorporates user interaction directly into the segmentation algorithm.

Many user-driven methods seek a balance between the level of intuition and user feedback that is useful for the reconstruction application. Successful methods tend to tightly integrate the reconstruction algorithm with the form of user interaction \cite{berger2017survey}.

The technique by Nan et al.~\cite{nan2010smartboxes} allows the user to approximately draw shape primitives comprised of axis-aligned boxes. These are compounded and repeated to form complex shapes, which are then iteratively fit to the point cloud. `Smartboxes' is geared towards the reconstruction of large urban environments exploiting regularities and symmetries.

With O-Snap, Arikan et al.~\cite{arikan2013snap} propose a shape-assisted interactive reconstruction system. 
The user's construction tools use best-fit plane segments recovered from the point cloud in a preprocessing step as work area. 
In a background optimization process, the user's corrections and constructions are constantly refitted to the point cloud. 
Edges and corners of polygons are snapped together if they are close enough. 
O-Snap reconstructs best-fit results based on a global optimization process. 
In contrast, our technique fundamentally aims at high-precision reconstruction of local point cloud features based on measured data.
However, in principle, O-Snap's continuous polygon optimization loop could be integrated into our approach if required by the application. 

Similar to our method, Steinlechner et al.~\cite{steinlechner2019adaptive} use shape detection on level-of-detail point data to locally extract support geometry. 
They apply a RANSAC-based method on the rendering data structure. 
In their approach, segmented planes only assist selection-related point interactions such as picking, brushing, or measurements. 

Lejemble et al.~\cite{lejemble2020persistence} find all stable planar features and relationship information through multi-scale analysis of the point cloud.
They give the user the ability to query for desired plane segments through brushing gestures. 
Lejemble et al. provide an integrated reconstruction workflow similar to ours. 
Conceivably, their global-segmentation based approach could be placed in the same system as our local ad-hoc reconstruction, combining their powerful segment browsing capabilities with our segment-assisted construction tooling.

Following the categorization provided by Berger et al. ~\cite{berger2017survey}, our approach does not require data-\emph{priors}, apart from the construction of an octree data structure that most renderers also use for level-of-detail rendering. Our approach applies to \emph{shape classes} of buildings indoor and outdoor, robustly deals with the \emph{artifacts} of \emph{missing data}, \emph{outliers}, and \emph{non-uniform sampling}, and outputs corners and edges. The artifact of \emph{misalignment} does not play a role in survey quality laser scans, where precise manual alignment is commonly performed beforehand.

\section{Incremental Region-Growing Point-Cloud Segmentation}
\label{sec:pointcloudsegmentation}

In this section we present a segmentation technique based on incremental seeded planar region growing. 
It works on a density-homogenized representation of a point cloud achieved through an octree-based adaptive resolution scheme. 
Automatic parameter estimation, allows for intuitive user-driven control of the algorithm. 
A graphical overview is shown in Figure \ref{fig:graphicalabstract}.

The notation $(\cdot ,\cdot ,\cdot ,\cdot )$ denotes a tuple of values. The boldfaced variable name $\mathbf{p} = (p_{x},p_{y},p_{z})$ stands for the three dimensional point $\mathbf{p}$ and its three components $p_{x}$, $p_{y}$ and $p_{z}$.

\subsection{Spatial Subdivision Scheme}
\label{subsec:octree}

Octrees, or equivalent spatial subdivision data structures, are commonly used for level of detail rendering and point cloud editing \cite{scheiblauer2011out}. 
In our out-of-core implementation, the leaf cells store references to point data, while the inner cells store references to subsampled representations of the contained cells. Apart from creating the octree, no further preprocessing on the point cloud is needed.
We use a simple cell indexing scheme similar to Yoder et al. \cite{yoder2006practical} to find adjacent cells and traverse cell neighborhoods. 

\subsection{Incremental Plane Regression}
\label{subsec:planeregression}

A prerequisite for our region growing method is the availability of an \emph{plane regression}. It is later used to define plane segments and incrementally updated given a user-defined seed point (Section \ref{subsec:initialseed}).

Let $R=(\mathbf{S},\mathbf{Sq},c,\mathbf{D})$ be an \emph{incremental plane regression} (Equation \ref{eq:incrementalplaneregression}), 

\kern-0.55em
\begin{align}
\label{eq:incrementalplaneregression}
&{\mathbf{S}}=\sum_{i=0}^{c}p_{i}\nonumber\\
&{\mathbf{Sq}}=\sum_{i=0}^{c}(p_{x_i}^{2},p_{y_i}^{2},p_{z_i}^{2})\\
&{\mathbf{D}}=\sum_{i=0}^{c}(p_{y_i}*p_{z_i},p_{x_i}*p_{z_i},p_{x_i}*p_{y_i})\nonumber
\end{align}

where $\mathbf{S}$ is the sum of point coordinates, $\mathbf{Sq}$ is the sum of squared point coordinates, $c$ is the point count and $\mathbf{D}$ is the vector of sums of products between different pairs of point coordinates. Let $\mathbf{p} = (p_{x},p_{y},p_{z})$ be a point, and $R'=(\mathbf{S'},\mathbf{Sq'},c',\mathbf{D'})=update(R,\mathbf{p})$ be the \emph{incrementally updated} planar regression produced by adding $\mathbf{p}$ to $R$. Then $R'$ is calculated as (Equation \ref{eq:incrementalupdate}):

\kern-0.55em
\begin{align}
\label{eq:incrementalupdate}
&{\mathbf{S}}'=\mathbf{S}+\mathbf{p}\nonumber\\
&{\mathbf{Sq}}'=\mathbf{Sq}+(p_{x}^{2},p_{y}^{2},p_{z}^{2})\\
&{c}'=c+1\nonumber\\
&{\mathbf{D}}'=\mathbf{D}+(p_{y}*p_{z},p_{x}*p_{z},p_{x}*p_{y})\nonumber
\end{align}

Given a plane regression $R=(\mathbf{S},\mathbf{Sq},c,\mathbf{D})$, we compute the associated \emph{covariance matrix} $C$. It represents the distribution of points in space, and later yields the parameters of the best-fit plane. Using the identity $\mathbf{\bar{p}}=\mathbf{S}/c$, the equation is (Equation \ref{eq:covmatrix}):

\kern-0.2em
\begin{equation} 
\label{eq:covmatrix}
\COV= 
\begin{bmatrix}
\SS_x-\AVG_x*\S_x & \CD_z-\AVG_x*\S_y & \CD_y-\AVG_x*\S_z \\ 
\CD_z-\AVG_x*\S_y & \SS_y-\AVG_y*\S_y & \CD_x-\AVG_y*\S_z \\ 
\CD_y-\AVG_x*\S_z & \CD_x-\AVG_y*\S_z & \SS_z-\AVG_z*\S_z
\end{bmatrix}
*\frac{1}{\CT - 1}
\end{equation}

The incremental formulation permits calculating the covariance matrix without having to recalculate the point centroid after every point addition. 
The actual set of included points does not need to be maintained and the memory and calculation overhead for adding a point to the regression is always constant.

Although our formulation is mathematically correct, real-world point clouds can exhibit very large point coordinates. 
These may lead to numerical problems, mainly due to the involved sum of squares. 
We shift all points by the coordinates of the first added point, such that the first point has the relative coordinate $(0,0,0)$. 
The covariance matrix can be \textit{rebased} accordingly after its computation. 
For this optional improvement, we use an equivalent implementation as published in Ponca~\cite{Ponca}. 
Our formulas can be found in Appendix \ref{sec:appendix_rebase}.

The evaluation in Section \ref{subsec:reconstructionaccuracy} shows that this improvement allows us to achieve an accuracy similar to standard approaches on real-world data sets, while maintaining the single-pass character of the approach.

\subsection{Plane Synthesis}
\label{subsec:planesynthesis}

Given the covariance matrix $C$, we synthesize the associated regression plane using the \emph{eigendecomposition} of $C$ into its principal components, i.e. finding the roots of the \emph{characteristic polynomial}

$det(C-\mathbf{\lambda} *I)=0$

to obtain the \emph{eigenvalues} $\lambda_{i},i\in\left \{0,1,2\right \}$, sorted by descending magnitude. Solving the specific eigenvalue equation

$(C-\mathbf{\lambda} *I)*\mathbf{v}=0$

\begin{sloppypar}
yield the three associated \emph{eigenvectors} ${\mathbf{v}_{i},i\in\left \{0,1,2\right \},\mathbf{v}_{i}\ne\mathbf{0}}$.
\end{sloppypar}

We obtain the unsigned plane normal as the eigenvector $\mathbf{v}_{2}$ associated with the smallest eigenvalue $\lambda_{2}$, tangential and bitangential directions as $\mathbf{v}_{0}$ and $\mathbf{v}_{1}$, and the plane origin as the last added point.

\subsection{Initial Seed}
\label{subsec:initialseed}

An incremental plane regression $R_{0}$ is initialized from a seed point $\mathbf{p}_{0}$ plus neighboring points. The seed point is chosen by the human operator via point picking. 
In case of an edge or corner seed, two or three plane regressions are initialized for the participating planes. 

In order to find the participating planes, we place a spherical \emph{seed region} of radius $r_{s}$ around $\mathbf{p}_{0}$. 
Within the seed region, seed points are chosen from the point cloud at a specific resolution, corresponding to octree cells at a certain level.
We suggest an adaptive resolution technique, which is further described in Section \ref{subsec:adaptiveresolution}. 
It ensures an appropriate subsampling rate for constant point density, even in the presence of severe anisotropy (e.g., floor around scanner).
The absolute subsampling rate, i.e., the \emph{desired point density} $d$, is chosen with our automatic parameter estimation technique, further described in Section \ref{subsec:parameters}.
We apply RANSAC plane fitting \cite{schnabel2007efficient} to find the initial planes, ranking them by inlier counts as stability measure.
A plane regression is initialized for each plane.
Optionally we discard the least stable planes in case of noisy scans using a variance-minimizing threshold~\cite{otsu1979threshold}. 
The combination of these systems gives control over the degree of detail, i.e., the size and complexity of the desired reconstruction results. This is independent of heterogeneity in point density or actual complexity of the scanned region.

\subsection{An Interactive Region Growing Loop}
\label{subsec:regiongrowingloop}

 A point's distance to a regression is its minimal Euclidean distance to the regression's associated plane. The decision to add a newly encountered point $\mathbf{p}$ to a regression $R$ is made if the point's distance $distance(R,\mathbf{p})$ to the regression is less than the \emph{plane threshold} $t_{p}$. 
 If true, we call the point an \emph{inlier point}. $t_{p}$ represents the maximum acceptable point distance and controls the precision of the reconstruction. 
 We refer to a regression together with its inlier points as a \emph{segment}.

\begin{figure}[t]
  \includegraphics[width=\linewidth]{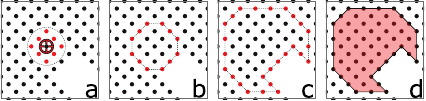}
  \caption{Interactive region growing loop. Steps (a) through (d) happen in sequence.}
  \label{fig:regiongrowing}
\end{figure}

Given initial seed regressions, the core reconstruction loop essentially consists of repeatedly adding neighboring points to the regressions if they lie within the plane threshold $t_{p}$. 
We thereby gradually expand outwards the set of \emph{border points} (inlier points that have unvisited points in their neighborhood). 
After the current octree cell is exhausted, and if any points were added to a regression, we communicate the current result using a \emph{progress callback} (Section \ref{subsec:progresscallback}). 
Then repeat the process for all bordering cells. 
In summary, the region growing algorithm is as follows:

\begin{enumerate}
  \item With increasing distance to the seed point, find unvisited point $\mathbf{p}$ closer to a regression $R$'s border than the \emph{search radius} $r$, and, if $distance(R,\mathbf{p})<\mathbf{t}_{p}$, $update(R,\mathbf{p})$. Repeat until all candidate points in the current cell have been visited. Update the average neighbor distance $d_{avg}$ with the actual distance between $\mathbf{p}$ and $R$'s border.
  \item Run the progress callback.
  \item If any point was added to any regression, repeat steps 1 through 3 for this cell's neighboring cells, with increasing distance to the seed point. Optionally, if $d_{avg}$ differs considerably from the desired density $d$, adjust the levels of neighboring cells according to our adaptive resolution scheme (Section \ref{subsec:adaptiveresolution}).
  \item The algorithm terminates after all candidate cells have been visited. The first few iterations of the algorithm are illustrated in Figure \ref{fig:regiongrowing}.
\end{enumerate}

\subsection{Progress Callback}
\label{subsec:progresscallback}

Leveraging incrementality, i.e. every intermediate result is also a valid final result, we return the current regression plane and \emph{regression polygon}, i.e. a flat polygon showing the outline of the border points projected onto the regression plane. 
We use a space carving technique equivalent to Alpha Shapes \cite{edelsbrunner1994three} to triangulate the regression polygon. 
We also preserve some useful per-segment statistics from the algorithm as output: point count, polygon area, and the \emph{regression variance} $\lambda_{2}$. $\lambda_{2}$ was calculated in Section \ref{subsec:planesynthesis}, and represents the mean squared error of the regression inliers along the plane's normal direction. 

If multiple regression planes exist in the current plane segmentation, we facilitate additional user interactions by finding stable edges and corners (Section \ref{subsec:edgesandcorners}).

\subsection{Adaptive Resolution}
\label{subsec:adaptiveresolution}

Point clouds are typically captured at different densities, depending on environmental factors such as size, distance to the scanner, and so on. 
Variations in point density have significant implications on performance and termination behavior of segmentation algorithms.
Parameters like the point search radius $r$ lose their meaning across regions with different densities. 
Globally downsampling the point cloud to a constant density is undesirable for reconstructing detailed structures, such as staircases, whose shapes would be lost. 
We propose an adaptive resolution approach. 
It dynamically adjusts cell levels during the region growing process, based on the observed point density $d_{avg}$ and the desired point density $d$. 

Octree cells correspond to regions in space, and \emph{parent} cells contain downsampled versions of its eight contained \emph{children} cells. The region growing algorithm places cells in a processing queue, in order of distance to the seed point. If $d_{avg}$ significantly differs from $d$, these cells are replaced in the queue by cells of higher or lower level, representing lower or higher sampling rates and point densities, respectively. 

A cell can simply be replaced by its children since they describe the same three dimensional space. 
Placing a cell's parent in the queue, however, is more involved. 
The parent comprises regions that have potentially already been processed, and should not be visited a second time. 
To avoid discarding information, we lock the current level. 
We completely finish processing the cells inside partially visited cells, and delay the transition to the parent level until this is complete. 
Our per-cell scheme is not optimal in speed since the reconstruction happens at a too fine detail level during the delay period. 
An alternative would incrementally remove already-processed areas from the regression and reconstruct them anew on a coarser level. 
The benefits of this additional optimization were negligible for our use case.

\begin{figure}[h]
  \includegraphics[width=\linewidth]{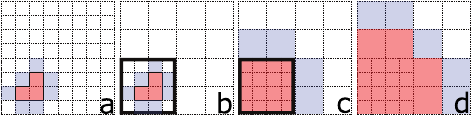}
  \caption{Adaptive resolution switch to a lower cell level. Steps (a) through (d) happen in sequence.}
  \label{fig:adaptiveresolution}
\end{figure}

In summary, the adaptive resolution algorithm consists of two cases:

\begin{figure*}
  \centering
    \includegraphics[width=\linewidth]{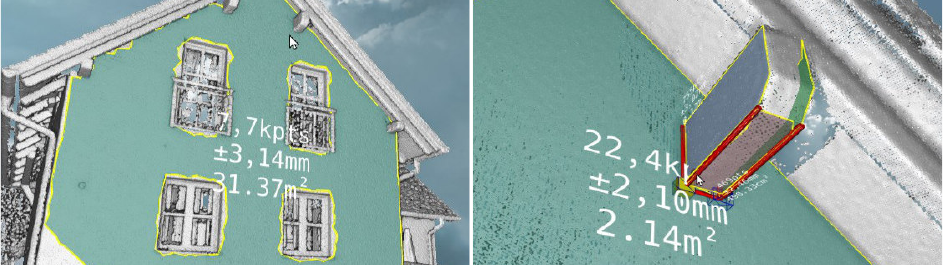}
  \caption{Zoom out (left) to obtain a planar regression of the entire house wall. Zoom in (right) to capture details in a local part of the wall.}
  \label{fig:segmentationcoarsevsfine}
\end{figure*}

\emph{Increase point density,} in case $d_{avg} \ll d$. Remove the current cell from the queue and replace it with its eight children. The cell level is now increased by 1. Continue processing the queue as usual. 

\emph{Decrease point density,} in case $d_{avg} \gg d$. Identify all cells inside the parents of completed cells or parents of cells in the queue. \emph{Mark} those cells. Continue processing the queue only for cells that are marked, until all marked cells have been visited. For all remaining cells in the queue, if $d_{avg} \gg d$ is still true, replace them with their parents. The cell level is now decreased by 1. Continue processing the queue as usual. This second case is illustrated in Figure \ref{fig:adaptiveresolution}. 

The thresholds for the level switch comparisons are free parameters. We used $d_{avg} < 0.5 * d$ for density increase, and $d_{avg} > 4.0 * d$ for density decrease, which worked reasonably well in our test setups. 
Our choice is based on the observation that planes are roughly split into four equal parts in the octree level transition. 
In the case of density increase, we err on the side of caution. A too-fine point sampling is preferable to a too-coarse one in terms of accuracy. 
The introduced sampling errors are further analyzed in Section \ref{subsec:reconstructionaccuracy}.

\subsection{Automatic Parameter Estimation}
\label{subsec:parameters}

Our incremental region growing point cloud segmentation depends on the following parameters. 
We derive \emph{intuitive controls} for the parameters, primarily based on the concept of moving the cursor and the camera closer to the site of reconstruction. 
This automatically increases the degree of reconstruction detail:

\begin{itemize}
    \item $\mathbf{p}_{0}$: seed point. $\mathbf{p}_{0}$ is chosen by picking a point from the point cloud using the cursor. 
    \item $r_{s}$: seed radius in [$m$]. $r_{s}$ is approximately 10\% of the screen width at the depth of $\mathbf{p}_{0}$. 
    \item $d$: desired point density (degree of detail) for the plane segmentation. Measured in [points per $m^{3}$]. $d$ is proportional to the projected size of a pixel around $\mathbf{p}_{0}$. Experimental values are between 10000 (closest to camera) and 100 (farthest away from camera), with a logarithmic falloff. 
    \item $r$: point neighbor search radius in [$m$]. $r$ is twice the estimated average point distance, depending on the point cloud's overall scale.
    \item $t_{p}$: plane distance threshold measured in [$m$]. $t_{p}$ remains as a free user parameter representing the desired precision of the plane regression.
\end{itemize}

Our objective is to provide human operators with direct and intuitive ways of controlling the algorithm. 
In particular, the initial seed selection (Section \ref{subsec:initialseed}) heavily determines the reconstruction and can be chosen to fit a user's particular needs. 
By moving the camera and zooming in, the users decide what constitutes a "good" or "bad" initialization. 
They are guaranteed to obtain the most salient planar features at the degree of detail they are currently viewing the point cloud at.
Figure \ref{fig:segmentationcoarsevsfine} gives an example.

\section{Assisted Geometry Construction Tools}
\label{sec:assistedgeometryconstructiontools}

In this section we present our methodology of robust edge and corner reconstruction (Section \ref{subsec:edgesandcorners}) and assemble feature-assisted modeling tools for interactive geometry reconstruction (Section \ref{subsec:modellingtools}).

\subsection{Edge and Corner Reconstruction}
\label{subsec:edgesandcorners}

\begin{figure}[h]
  \centering
  \includegraphics[width=0.475\textwidth]{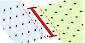}
  \caption{Stable edge support (red) between two plane segments (green and blue) where both segments have inlier border points (inside dotted lines).}
  \label{fig:supportline}
\end{figure}

\begin{figure*}
  \begin{subfigure}[]{\linewidth}
    \centering
    \includegraphics[width=\linewidth]{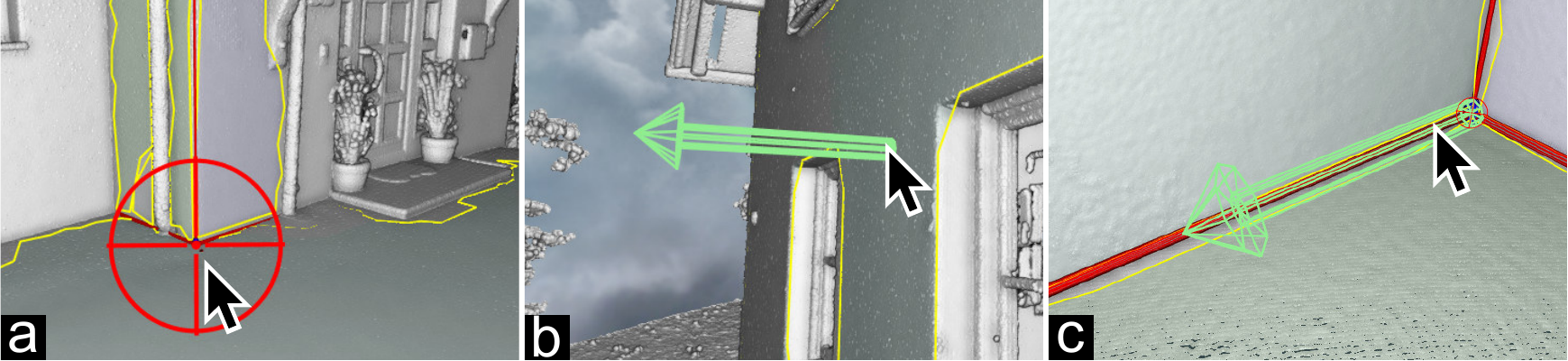}
  \end{subfigure}
  \caption{Shape-assisted interactions. (a) corner snapping. (b) selecting plane normal. (c) selecting plane tangent.}
  \label{fig:snapping}
\end{figure*}

\begin{figure*}[b]
  \centering
  \includegraphics[width=\linewidth]{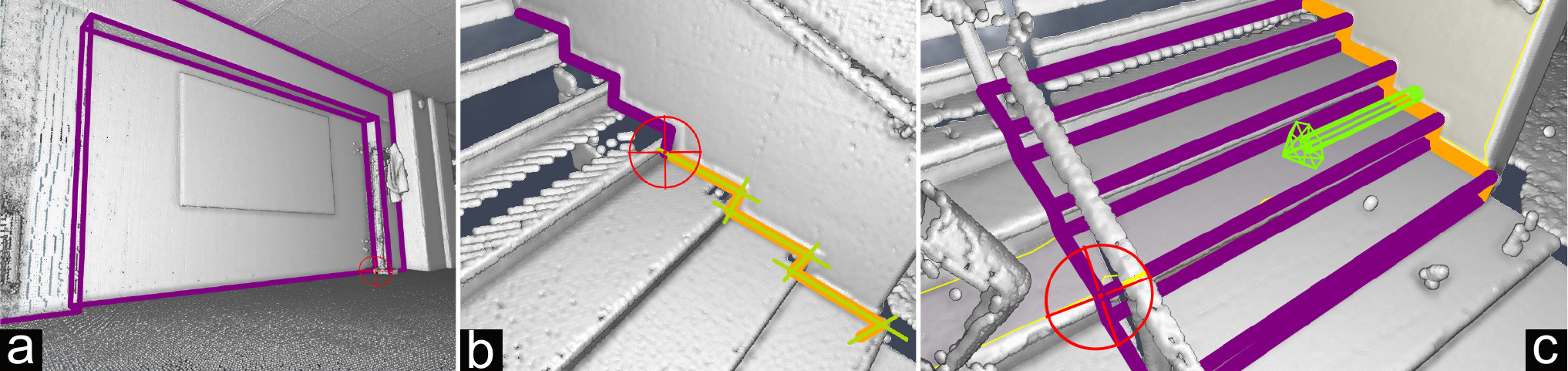}
  \caption{Plane-assisted CAD tools. (a) polygon construction. (b) polygon movement. (c) polygon extrusion.}
  \label{fig:tools}
\end{figure*}

We automatically reconstruct the \emph{maximum likelihood} edges and corners as intersections between regressions. Given a region growing result, all two- and threefold selections of regressions are intersected if:

\begin{itemize}
\item the regression planes are not parallel, 
\item the regression polygons have non-vanishing areas, and
\item after intersecting the regression planes, all regression polygons have vertices in proximity to the intersection line/point.
\end{itemize}

The resulting set of intersections has elements of two categories, which receive stability measures called \emph{supports}:

\emph{Corners.} The intersection between three planes results in a corner. The viable corner's support is comprised of the three regression variances.

\emph{Edges.} The intersection between two planes results in an edge. For a viable edge, the participating segments have border points closer than $2*\mathbf{r}$ to the intersection line. Projecting these border points onto the line gives a 1D distribution. The 95th percentile of it is the \emph{support range} and produces the stable start- and endpoints of the line segment describing the edge (illustrated in Figure \ref{fig:supportline}).

\subsection{Feature-Assisted Modeling Tools}
\label{subsec:modellingtools}

Based on our experience with experts and novice users, we identify two principal modes of shape assistance for reconstruction workflows: \emph{Cursor snapping} and \emph{direction finding}.

\emph{Cursor Snapping} enables the human to point out dominant structures: maximum likelihood edges and corners (Figure \ref{fig:snapping}(a)).

\emph{Direction finding} allows the user to select directions in three dimensional space. 
The human operator chooses a direction by simply pointing at a plane to select the plane normal. Alternatively, pointing at an edge or a corner selects one of the planes' tangents through subtle pointer movement. 
This interaction is shown in Figure \ref{fig:snapping}(b) and \ref{fig:snapping}(c).
We aim at offering easy-to-understand tools for the user by only adhering to simple pointing-based selection techniques.

Leveraging the described support mechanisms, we implement an essential set of \emph{feature-assisted CAD tools} (shown in Figure \ref{fig:tools}):

\begin{itemize}
    \item \emph{Free polygon construction} assisted by pointer snapping (Figure \ref{fig:tools}(a)).
    \item \emph{Moving and copying polygons} assisted by pointer snapping and optional direction constraining (Figure \ref{fig:tools}(b)).
    \item \emph{Polygon extrusion} constrained along a fixed direction, further assisted by polygon snapping (Figure \ref{fig:tools}(c)).
\end{itemize}

\section{Evaluation}
\label{sec:evaluation}

We assess our technique in three different ways by determining the:

\begin{itemize}
    \item \emph{reconstruction accuracy} of our out-of-core incremental region growing algorithm with synthetic test cases,
    \item \emph{effectiveness} of our workflow and interactions with example use cases and time measurements, and
    \item \emph{applicability in real-world scenarios} through interviews with surveying experts based on a version of our framework implemented in their application.
\end{itemize}

The performance of the region growing method is crucial for interactivity. While our incremental region growing formulation (Section \ref{subsec:planeregression}) is independent of the point cloud size, interactions depend on the chosen level of subsampling (Section \ref{subsec:reconstructionaccuracy}). On consumer-level hardware (Intel i5-4690K, Nvidia GTX 1080) the extraction of geometrical features and region growing typically took between 0.1 and 1 seconds depending on plane size and never slowed down interaction, especially since the first results appear immediately and refinement happens until full convergence.

\subsection{Reconstruction Accuracy}
\label{subsec:reconstructionaccuracy}

We evaluate two kinds of reconstruction accuracies for our proposed technique: The \emph{absolute reconstruction error} measured as a function of input noise to the precision of reconstructed features, and the \emph{effect of subsampling} as the loss of precision if lower point densities are chosen. 

The absolute reconstruction error (as Root Mean Square Error of inlier points to their regression planes) is evaluated in comparison to the well-established robust RANSAC plane detection algorithm by Schnabel et al. \cite{schnabel2007efficient}. 
We chose this particular algorithm as baseline because our collaboration partners work with it regularly and deem it sufficiently accurate for real-world scenarios. 
Implementations are freely included in open-source point cloud processing tools such as CloudCompare \cite{cloudcompare}.
We create a synthetic scenario with known input noise, and process the input by both algorithms in a controlled fashion.
Our synthetic scenario is generated using a simulated laser scanner that is placed inside a room-sized box. 
It scans its surroundings in regular patterns, introducing depth measurement errors ranging between one millimeter and a few centimeters. 
The measurement error is simulated in a realistic way, i.e., the virtual laser scanner produces angular and depth measurement errors similar to a real one, leading to realistic anisotropy and error-to-distance behaviors.
All tests produce between $10^{5}$ and $10^{6}$ points, which is small enough to fit into the core on typical hardware, and are repeated at least 200 times. 
Pre-defined user inputs are simulated to reconstruct the corners of the synthetic room. 
The accuracy of them is then compared to that of the RANSAC plane detection and the subsequent recovery of the same corners. 

\begin{figure}[t]
  \centering
  \includegraphics[width=0.48\textwidth]{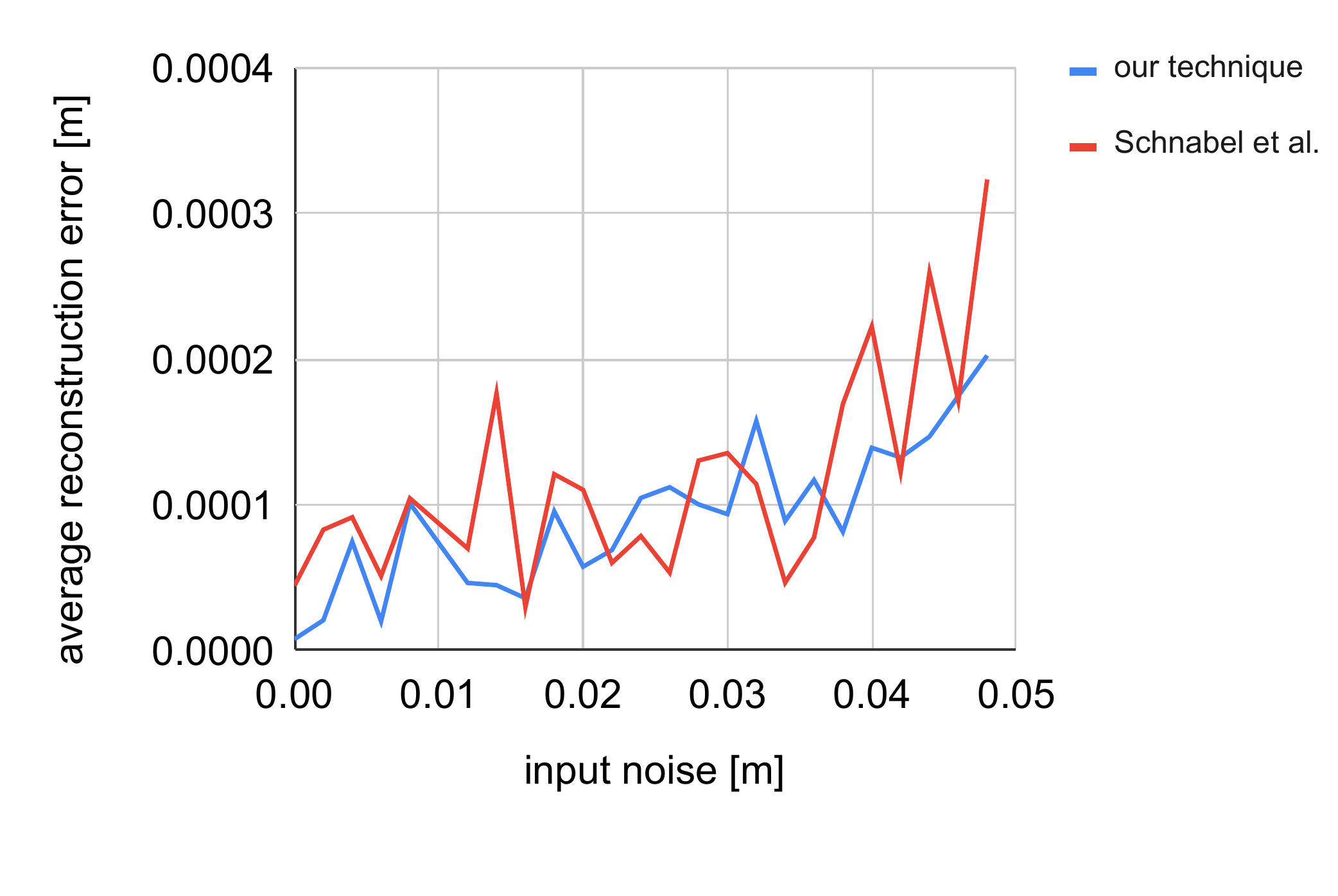}
  \caption{Input noise vs. reconstruction error at maximum detail. Lower is better. Our proposed system in blue, RANSAC plane detection by Schnabel et al. \cite{schnabel2007efficient} in red. Synthetic data set.} 
  \label{fig:reconstructionerror}
\end{figure}

The evaluation results are illustrated in Figure \ref{fig:reconstructionerror}. 
The graph clearly shows that our incremental region growing algorithm is roughly in line with the established technique. 
Both approaches produce reconstructions approximately an order of magnitude more precise to the ground truth than the scale of the point cloud noise. 
This evaluation shows that our incremental reconstruction approach suffers no compromise in accuracy if applied to the base case (in-core point cloud and no user input).

\begin{figure}[t]
  \centering
  \includegraphics[width=0.48\textwidth]{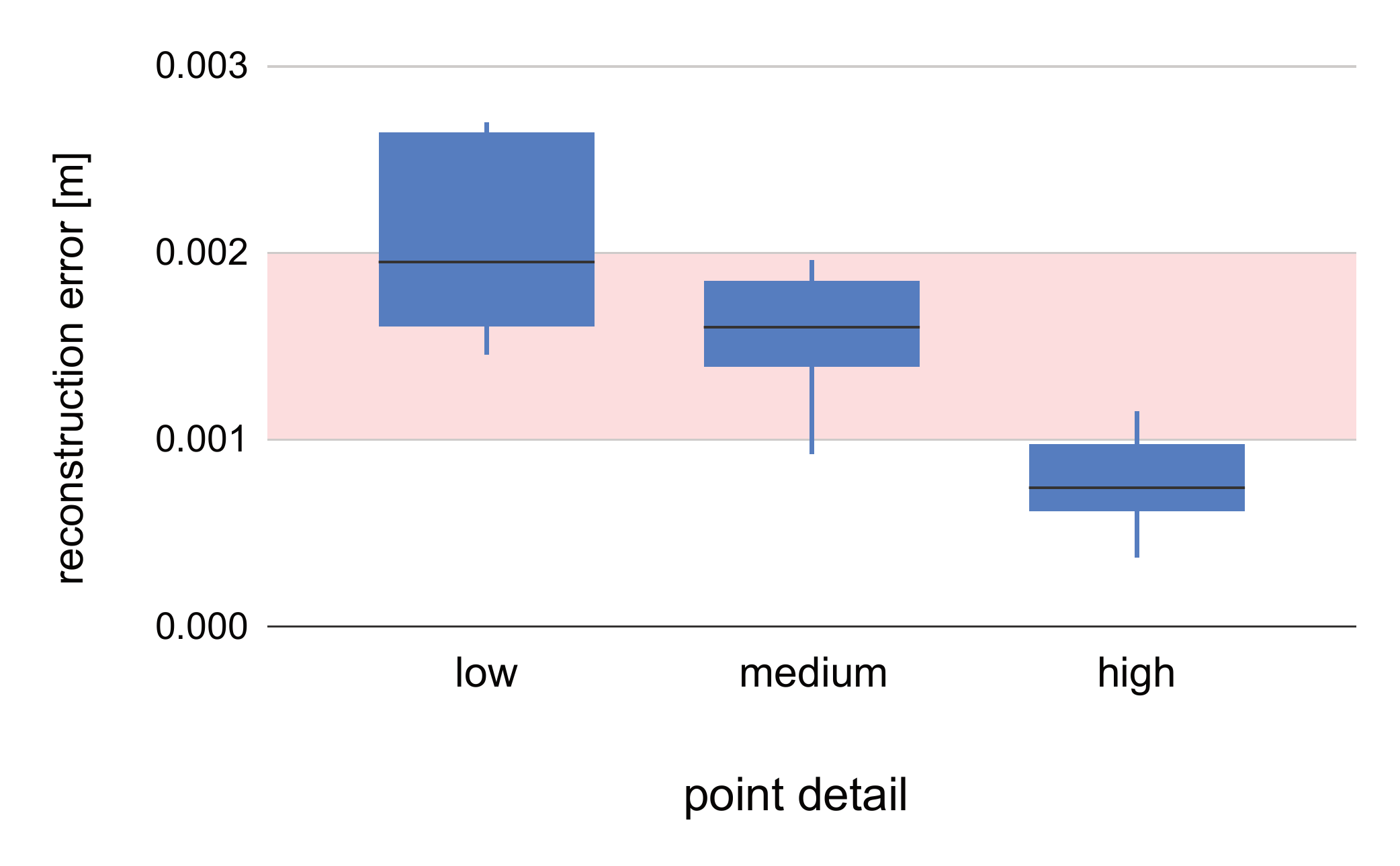}
  \caption{Degree of point detail vs. reconstruction error. Lower is better. Red area shows the approximate scanner noise. Our detail levels are: \emph{low} ($>$2cm resolution), \emph{medium} (1mm-2cm), and \emph{high} ($<$1mm). Real-world data set, dimensions: approx. 7m x 5m x 2.5m.}
  \label{fig:reconstructionerrorvsdetail}
\end{figure}

The use of subsampling in our level-of-detail region growing leads to a certain loss of precision. 
To evaluate this in a real-world context, we chose a high-resolution (1 mm - 2 mm point distance) laser scan of a building interior and selected a well-captured corner. 
As opposed to the previous analysis, we now recover a feature of a scan with a known, fixed noise level at different subsampling rates.
We repeatedly reconstructed the corner with our plane segmentation at different point densities (segmentation parameter $d$). 
This measures the impact of our adaptive resolution scheme on the reconstruction accuracy.
The point densities are categorized into \emph{low detail} cases with more than \mbox{2 cm} between points, \emph{medium detail} cases ranging from \mbox{1 mm} to \mbox{2 cm} between points, and \emph{high detail} with less than \mbox{1 mm} between scan points. 
The values correspond to detail levels commonly referenced in the domain of urban surveying. 

Figure \ref{fig:reconstructionerrorvsdetail} shows the results. 
The input scanner noise is highlighted as red area and lies between \mbox{1 mm} and \mbox{2 mm}. 
In the low detail reconstruction, the output error lies between \mbox{1.5 mm} and \mbox{3 mm}.
In the medium detail reconstruction, the output error is roughly equal to the input noise. 
In the high detail reconstruction, the output error lies below the input noise, at approximately \mbox{0.5 mm} to \mbox{1 mm}.
These results show that the loss of precision introduced by subsampling in our technique stays within tolerable bounds.
The domain experts found these accuracy ranges appropriate for typical reconstruction workflows that involve multiple levels of detail.

\subsection{Example Use Cases}
\label{subsec:exampleusecases}

To demonstrate the capabilities of our technique, we reconstructed example scenarios from real-world use cases. 
The examples' scopes are similar to that of common tasks in the domain of as-built surveying. 
The point clouds were supplied by our collaboration partners and were not cleaned or downsampled beforehand. 
These data sets range from approx. 10 million to over 1 billion (and, in real applications, up to 10+ billion) points. 

To put this number into perspective, we found it only possible to load up to 50 million points into comparable interactive systems like Arikan et al. ~\cite{arikan2013snap} and Lejemble et al.~\cite{lejemble2020persistence}. Beyond 50 million points, memory and runtime constraints of global pre-processing proved prohibitive.
Our local technique, with adaptive subsampling, keeps memory and runtime proportional to the size and detail of the reconstruction, rather than the size of the point cloud.

All tests were done on average consumer hardware, i.e., Intel i5-4690K, Nvidia GTX 1080.

\begin{figure*}[t]
  \centering
  \includegraphics[width=\linewidth]{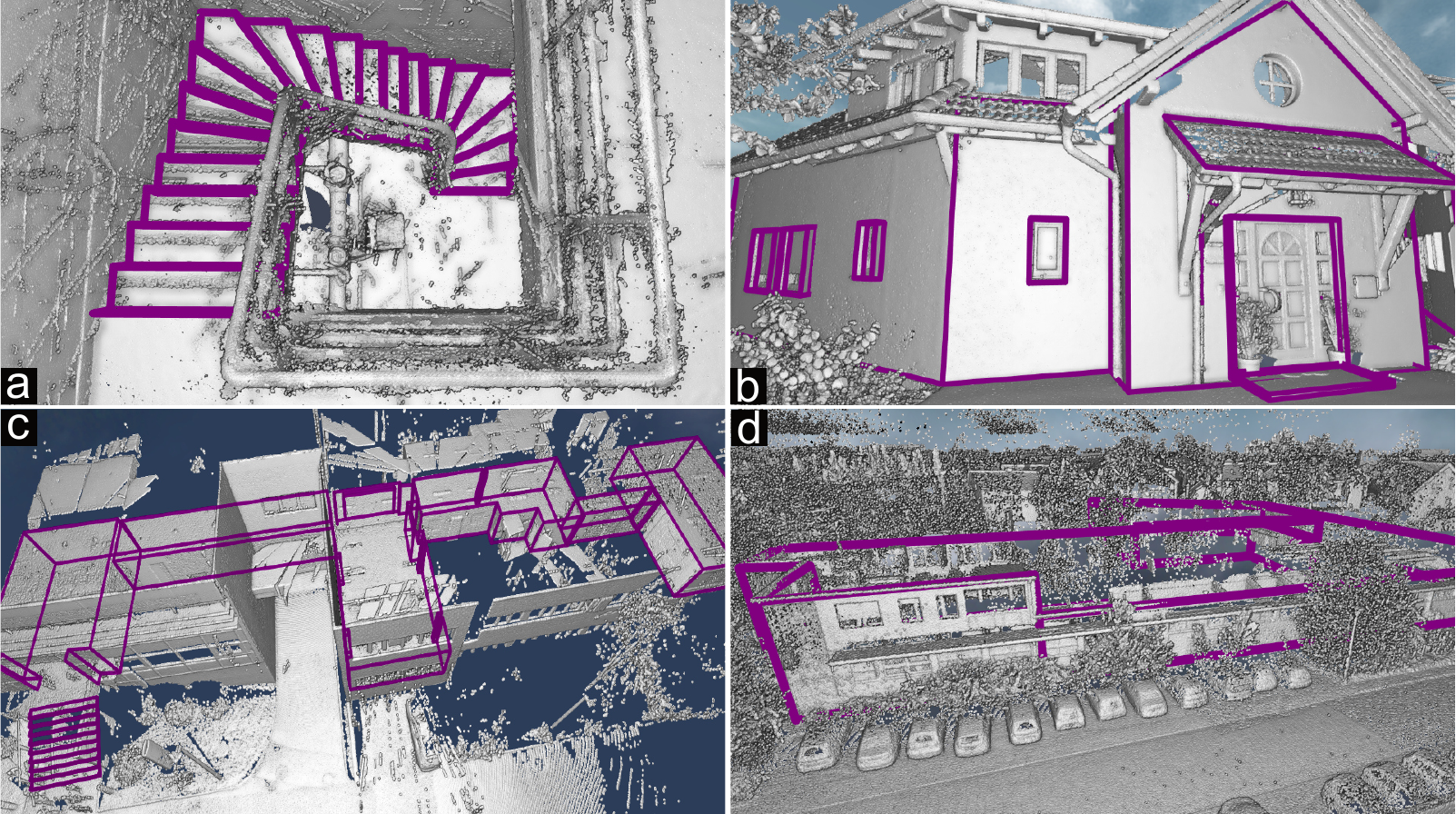}
  \caption{Reconstructed scenarios from Table \ref{tab:scenarios}. (a) Indoor spiral staircase. (b) House facade. (c) Indoor hull of an office building that was scanned both from inside and outside. Traditionally prone to severe oversegmentation, this frequent scenario was straightforwardly reconstructed using our method by moving the camera through the rooms and clicking on the corners. (d) Factory layout. The facade is severely obscured by trees, cars, and moving people, which are common artifacts in mobile mapping. This scenario was reconstructed using our method by clicking on the dominant planes from a zoomed-out view, and ad-hoc filtering the noise.}
  \label{fig:scenarios}
\end{figure*}

The first scenario, `office interior', was reconstructed by an expert user.
They are an industrial provider of surveying services, with more than five years of experience and familiar with the traditional reconstruction workflow. 
The expert is not affiliated with this paper.
The expert reconstructed the inside layout of an office building, using our demonstrator GUI.
It uses our interactive technique. They compared it with an expert tool that is being employed by our collaboration partners and has limited interaction assistance. 
The expert tool uses RANSAC-based global plane segmentation and allows the user to select and intersect planes to find edges. 
The user ended up reconstructing approximately 92 polygons. 
The result is visible in Figure \ref{fig:scenarios}(c). 
It took them approximately 20 minutes with our technique. 
Using the comparison tool with limited shape assistance, they required approximately 105 minutes. 
The main reason for this longer time is that the user had to run the global RANSAC segmentation repeatedly with different parameters. 
They were unable to find one set of parameters that was appropriate for the dense parts as well as the sparse parts of the point cloud. 
The user did not have this problem with our technique, since our instant feedback interaction allowed them to find locally appropriate segmentation parameters on the fly. 

The remaining example reconstruction scenarios are shown in Table \ref{tab:scenarios}, with screenshots in Figure \ref{fig:scenarios} and descriptions in the caption. 
Scenarios were selected to be representative of various real-world architectural disciplines: building layouts, building facade reconstruction, outdoor and indoor details, and varying sampling rates and occlusions/noise. 
The selection of three plane segments to identify the corner point at their intersection has been an interaction that was used particularly often, especially in cases of scanning shadows due to occlusion.
Whenever the users encountered a hole in the point cloud, they usually decided to extrapolate surrounding plane segments, or guess the missed shape from a different, symmetrical part of the building. 
This showcases the value of human understanding and contextual awareness to infer shapes from missing point data. 

The reconstruction example scenarios resulted in approximately 20-40 polygons and were completed in a matter of minutes. 
Users subjectively estimated the they would need more than an hour for similar reconstructions without shape assistance. 
They could navigate towards their reconstruction targets, and then invoke our on-demand plane segmentation. 
They quickly found reconstruction parameters appropriate for the current local region and reconstruction goal, regardless of the point cloud size and complexity.

\subsection{Expert Interviews}
\label{subsec:expertinterviews}
We conducted interviews with users who have varying degrees of experience: two domain experts who work with point clouds professionally, three intermediate users who had some point cloud experience, and two novice users who had no prior experience in this domain. 
The two domain experts work in terrestrial and building surveying, with more than three years of experience. 
The intermediate users apply point cloud tools regularly (at least two years of experience), but are not familiar with high-precision reconstruction. 
The novice users have no professional experience with point clouds or reconstruction.
None of the respondents are co-authors of this paper.
We asked them to complete minimal reconstruction tasks, answer usability questions, and give their subjective opinions on our technique. 

\begin{table}[t]
\scalebox{0.92}{
\begin{tabular}{||c c c c||} 
 \hline
 scenario & \#points & polygons & time taken \\ [0.5ex] 
 \hline\hline
 spiral staircase (\ref{fig:scenarios}a) & $\sim$15 million & 45 & $\sim$15 min. \\ 
 \hline
 house facade (\ref{fig:scenarios}b) & $\sim$50 million & 49 & $\sim$12 min. \\ 
 \hline 
 office interior (\ref{fig:scenarios}c) & $\sim$150 million & 92 & $\sim$20 min. \\ 
 \hline
 factory layout (\ref{fig:scenarios}d) & $\sim$1 billion & 22 & $\sim$8 min. \\ [1ex] 
 \hline
\end{tabular}
}
 \caption{Example scenarios reconstructed with our technique. Screenshots are shown in Figure \ref{fig:scenarios}.}
 \label{tab:scenarios}
\end{table}

The domain experts are accustomed to a workflow that involves manually drawing a polygon on an orthographic projection of a part of the point cloud. 
This is supported by coarse global pre-segmentation and direct point picking. 
Common usability issues of this workflow stem from the abrupt viewpoint switches between different orthographic depictions and the high cognitive load of selecting and repairing the output of global segmentation.

All respondents reported a strong feeling of presence in the data set, always being aware which part of the point cloud they were currently viewing. 
They zoomed the camera in and out of the point sets to switch between coarse and fine-grained reconstruction parameters. 
Consequently, they alternated between overview and detail views of the data on their screens. 
Our automatic parameter estimation naturally supports such focus-and-context interaction types.

One task was to reconstruct a rectangular wall on a building facade, of which one corner was missing due to a scan shadow. 
The intermediate and expert users intuitively selected surrounding planes and found the missing corner as the intersection point. 
All participants stated that they were confident in the interaction and that they felt the resulting geometry was identical to what they had intended. 
In the previous workflow, this task is more difficult. 
The automatic plane pre-segmentation needs to be adjusted to specific parameters to correctly select the participating planes for intersection. 

Another task was to reconstruct the volume of a window reveal, of which the bottom side was missing. 
The participants applied the extrusion tool to construct the volume starting from the top side and snapping to the bottom. 
The participants positively commented on the simplicity of the interaction, being able to complete it from a single viewpoint. 
In the previous workflow, this task requires multiple viewpoint switches to capture the entire three-dimensional volume. 

Ultimately, the novice and intermediate users found our approach to be a novel and interesting way of interacting with point clouds. 
This was evident from most of them staying to explore and reconstruct more point sets after the interview was over. 
The expert users welcomed our interaction techniques for being able to handle common tasks with ease that are otherwise cumbersome without feature support. 
An example isthe ability to select plane normals or tangential directions with a single click. 
They found our ability to interact with point clouds of arbitrary size without compromising reconstruction accuracy an essential asset for modern high-resolution laser scanners.

\subsection{Real-world Application}

To indicate the usability of our approach, we briefly touch on usage of our interaction system in real-world applications. 
We implemented our technique in an out-of-core octree and point cloud rendering system, which can handle point clouds of arbitrary size, in our experiments up to tens of billions of points. 
This exemplifies how our approach can be integrated in existing point cloud tooling.

The project partners and domain experts in urban surveying at rmDATA use our system in their commercial product \mbox{rmDATA 3DWorx}~\cite{rmdata3dworx}. 
They deploy it to expert users in the domains of as-built surveying and urban reconstruction. 
A domain expert independently measured lengths on buildings from which we have laser scans. 
Our reconstructed measurements agree with their precise surveying, showing \mbox{0.35 mm} of difference on average on a length of \mbox{1 m}.

\subsection{Comparison with Interactive Tools}

For completeness, we briefly compare the reconstruction experience using our tool compared to similar published work, O-Snap (Arikan et al.~\cite{arikan2013snap}) and Lejemble et al.~\cite{lejemble2020persistence}, applied to their publicly available demonstration data sets.

The clear advantage of O-Snap lies in a fast reconstruction as long as the data is available.
As noted in their evaluation, reconstruction takes longer in regions where building parts are not included in the original data.
Furthermore, their global automatic optimization focuses on creating a watertight mesh, which makes the precision of the reconstructed geometry difficult to comprehend.

Lejemble et al.'s technique is more similar to ours in terms of user interaction. 
It is also unable to use missing regions for modeling, but their effective segment-browsing toolset notably speeds up the process. 
Their tooling could be adapted to work within our local reconstruction loop. 

In summary, all three approaches provided adequate ways to obtain the desired geometry, but each one is fundamentally motivated by different reconstruction scenarios and requirements. 
In comparison, our technique took approx. 5-10 minutes longer than in the compared works. 
It, however, had a clear focus on completing reconstructions where others failed, and on modeling the geometry as accurately as possible.

\section{Discussion and Future Work}
\label{sec:discussion}

Our interactive and localized approach to point cloud reconstruction makes for lightweight interactions.
These are intuitively clear to humans and produce little cognitive load. 
Further, they interface directly with a point cloud.
All of these are valuable properties. 
Novice and expert users quickly adapted to our technique due to its intuitive nature and reported gaining new insight into the structure of point clouds they interacted with.

Currently, in our technique, shapes other than plane segments are not considered. 
An interesting research direction is the integration of other support geometries, such as parametric surfaces, or the automatic detection of different kinds of edges~\cite{hackel2016contour},~\cite{himeur2021pcednet} or parametric curves~\cite{wang2020pie}.
It remains to be investigated whether they could be assembled into sensible user interactions outside of highly specialized contexts.

Our feature-assisted CAD tooling requires no specialized point cloud infrastructure and can easily be integrated with existing interaction systems, such as O-Snap~\cite{arikan2013snap}. 
As future work, a mesh optimization loop similar to O-Snap's could be used to evolve our loose collection of reconstructed polygons into watertight meshes on-the-fly.

\section{Conclusion}
\label{sec:conclusion}

In this paper we present an interaction framework for feature-assisted point cloud geometry reconstruction using on-demand planar region growing.
The core contributions are a density-normalized out-of-core point cloud reconstruction technique, an on-demand plane-segmentation interaction framework, and a feature-assisted high-precision geometry reconstruction workflow.
Our evaluation shows that we achieve the stated goal of providing reconstruction capabilities at arbitrary precision and level of detail on large out-of-core point clouds without preprocessing.

\appendix
\appendix

\section{Rebasing Incremental Plane Regression}
\label{sec:appendix_rebase}

\begin{sloppypar}
Given an incremental plane regression $R=(\mathbf{S},\mathbf{Sq},c,\mathbf{D})$ as defined in Section \ref{subsec:planeregression}, and a new \textit{reference point} $\mathbf{r}$, with the previous reference point denoted as $\mathbf{r_{0}}$. 
Let ${\mathbf{d}=\mathbf{r}_{0}-\mathbf{r}}$, we calculate the \textit{rebased incremental plane regression} ${R^{R}=(\mathbf{S}^{R},\mathbf{Sq}^{R},c,\mathbf{D}^{R})}$ which re-centers the synthesized values around $\mathbf{r}$ using the following Equations \ref{eq:appendix_rebase}:
\end{sloppypar}

\small
\begin{align}
\label{eq:appendix_rebase}
&{\mathbf{S}}^{R}=\mathbf{S}+c*\mathbf{d}\nonumber\\
&{\mathbf{Sq}}^{R}=\mathbf{Sq}+2*(d_{x}*S_{x},d_{y}*S_{y},d_{z}*S_{z})+c*(d_{x}^{2},d_{y}^{2},d_{z}^{2})\nonumber\\
&v_{x}=d_{x}*S_{y}+d_{y}*S_{x}+c*d_{x}*d_{y}\\
&v_{y}=d_{x}*S_{z}+d_{z}*S_{x}+c*d_{x}*d_{z}\nonumber\\
&v_{z}=d_{y}*S_{z}+d_{z}*S_{y}+c*d_{y}*d_{z}\nonumber\\
&{\mathbf{D}}^{R}=\mathbf{D}+(v_{x},v_{y},v_{z})\nonumber
\end{align}
\normalsize



\section{Conflicts of interests/competing interests}
The authors have no competing financial or non-financial interests to disclose.

\section{Acknowledgment of Funding}
This work at VRVis is funded by BMK, BMDW, Styria, SFG, Tyrol and Vienna Business Agency in the scope of COMET-Competence Centers for Excellent Technologies (879730) which is managed by FFG.
Part of the work has been funded by the Vienna Science and Technology Fund (WWTF) and the City of Vienna (Grant ID: 10.47379 / ICT22055).


\bibliographystyle{cas-model2-names}
\bibliography{paper}


\end{document}